\documentclass[preprint,showkeys,superscriptaddress,amsmath,amssymb]{revtex4-1}

\usepackage{newtxtext,newtxmath}
\usepackage[T1]{fontenc}
\usepackage[pdftex]{graphicx}
\usepackage{bm}
\usepackage{textcomp}
\usepackage{braket}
\usepackage{siunitx}
\usepackage{enumitem}
\usepackage{caption}

\DeclareCaptionLabelSeparator{bar}{ $|$ }
\usepackage[colorlinks=true,
linkcolor=magenta,
citecolor=blue,
pdfauthor={Ehsan Arbabi},
]{hyperref}

\usepackage{times}
\usepackage{dcolumn}
\usepackage{amsmath}
\usepackage[dvipsnames]{xcolor}
\usepackage[labelsep=space]{caption}
\usepackage{setspace}
\usepackage{amssymb}
\usepackage[version=3]{mhchem}

\begin{document}

\renewcommand{\captionfont}{\small}

\title{Full Stokes imaging polarimetry using dielectric metasurfaces}

\author{Ehsan Arbabi}
\affiliation{T. J. Watson Laboratory of Applied Physics, California Institute of Technology, 1200 E. California Blvd., Pasadena, CA 91125, USA}
\author{Seyedeh Mahsa Kamali}
\affiliation{T. J. Watson Laboratory of Applied Physics, California Institute of Technology, 1200 E. California Blvd., Pasadena, CA 91125, USA}
\author{Amir Arbabi}
\affiliation{Department of Electrical and Computer Engineering, University of Massachusetts Amherst, 151 Holdsworth Way, Amherst, MA 01003, USA}
\author{Andrei Faraon}
\email{Corresponding author: A.F.: faraon@caltech.edu}
\affiliation{T. J. Watson Laboratory of Applied Physics, California Institute of Technology, 1200 E. California Blvd., Pasadena, CA 91125, USA}
\maketitle

\clearpage

\textbf{Polarization is a degree of freedom of light carrying important information that is usually absent in intensity and spectral content. Imaging polarimetry is the process of determining the polarization state of light, either partially or fully, over an extended scene. It has found several applications in various fields, from remote sensing to biology. Among different devices for imaging polarimetry, division of focal plane polarization cameras (DoFP-PCs) are more compact, less complicated, and less expensive. In general, DoFP-PCs are based on an array of polarization filters in the focal plane. Here we demonstrate a new principle and design for DoFP-PCs based on dielectric metasurfaces with the ability to control polarization and phase. Instead of polarization filtering, the method is based on splitting and focusing light in three different polarization bases. Therefore, it enables full-Stokes characterization of the state of polarization, and overcomes the 50$\%$ theoretical efficiency limit of the polarization-filter-based DoFP-PCs.}

\renewcommand{\figurename}{}
\renewcommand{\thefigure}{\noindent\textbf{Figure} \textbf{\arabic{figure}}}

Polarimetric imaging is the measurement of the polarization state of light over a scene of interest. While spectral and hyperspectral imaging techniques provide information about the molecular and material composition of a scene~\cite{Chang2003Hyperspectral,Sun2010Hyperspectral}, polarimetric imaging contains information about the shape and texture of reflecting surfaces, the orientation of light emitters, or the optical activity of various materials~\cite{Tyo2006ApplOpt,Garcia2015OptExpress}. This additional information has led to many applications for imaging polarimetry ranging from astronomy and remote sensing to marine biology and medicine~\cite{Tyo2006ApplOpt,Coffeen1979JOSA,Walraven1981SPIEOptEng,Egan1991ApplOpt,
Liu2012JBioOpt,Roberts2014ProcIEEE,Charanya2014SPIEJBioOpt,Garcia2017Optica}. Therefore, several methods have been developed over the past decades to enable mapping of the polarization state over an extended scene~\cite{Johnson1974AMERDC,Chin-Bing1976DTIC,Solomon1981ApplOpt,Pezzaniti1995OptEng,
Nordin1999JOSAA,Guo2000ApplOpt,Gruev2006IEEESCS,Garcia2017Optica}.

Generally, polarimetric imaging techniques can be categorized into three groups: division of amplitude, division of aperture, and division of focal plane~\cite{Tyo2006ApplOpt}. All of these techniques are based on measuring the intensity in different polarization bases and using them to estimate the full Stokes vector or a part of it. Among various systems, DoFP-PCs are less expensive, more compact, and require less complicated optics~\cite{Nordin1999JOSAA,Guo2000ApplOpt,Gruev2006IEEESCS}. In addition, they require much less effort for registering images of different polarizations as the registration is automatically achieved in the fabrication of the polarization sensitive image sensors. The advances in micro/nano-fabrication have increased the quality of DoFP-PCs and reduced their fabrication costs, making them commercially available. DoFP-PCs either use a birefringent crystal to split polarizations~\cite{Rust1995Patent,Andreou2002IEEESensJour}, or thin-film~\cite{Guo2000ApplOpt,Gruev2007OptExp} or wire-grid~\cite{Nordin1999JOSAA,Gruev2010OptExp,Garcia2017Optica} polarization filters. To enable the measurement of degree of circular polarization, form-birefringent quarter waveplates were integrated with linear polarizers in the mid-IR~\cite{Deguzman2001ApplOpt}. Recently, liquid crystal retarders have been integrated with linear polarization filters to enable full Stokes polarimetric imaging by implementing circular~\cite{Myhre2012OptExpress} and elliptical polarization filters~\cite{Hsu2014OptExpress,Hsu2015OptExpress}. An issue with the previously demonstrated DoFP-PCs is that they all have a theoretical efficiency limit of 50$\%$ due to using polarization filters~\cite{Tyo2006ApplOpt}, or spatially blocking half of the aperture~\cite{Andreou2002IEEESensJour}.

Optical metasurfaces are a category of nano-fabricated diffractive optical elements comprised of nano-scatterers on a surface~\cite{Kuznetsov2016Science,Staude2017NatPhoton,Hsiao2017SmallMeth,Astilean1998OptLett,
Lalanne1998OptLett,Fattal2010NatPhoton,Lu2010OptExp,Vo2014IEEEPhotonTechLett,
Lin2014Science,Arbabi2015NatCommun,Khorasaninejad2016Science,Yin2017LighSciAppl,
Schlickriede2017AdvMat,Komar2017ApplPhysLett,Paniagua2017arXiv,Zhan2017SciRep} that are judiciously designed to control the wavefront. They have enabled high-efficiency phase and polarization control with large gradients~\cite{Astilean1998OptLett,Lalanne1998OptLett,Arbabi2015NatCommun,
Arbabi2015NatNano,Khorasaninejad2016Science,Paniagua2017arXiv}. In addition, their compatibility with conventional microfabrication techniques allows for their integration into optical metasystems~\cite{Arbabi2016NatCommun,Arbabi2017NatPhoton,Groever2017NanoLett,Arbabi2017arXiv}.

Metasurfaces have previously been used for polarimetry~\cite{Ellenbogen2012NanoLett,Dandan2015OptExp,Khorasaninejad2015Optica,Chen2016Nanotechnology,Mueller2016Optica,Ding2017ACSPhotonics}, but not polarimetric imaging. An important capability of high contrast dielectric metasurfaces is the simultaneous control of polarization and phase~\cite{Arbabi2015NatNano}. Here, we use this capability to demonstrate a dielectric metasurface mask for DoFP-PCs with the ability to fully measure the Stokes parameters, including the degree of circular polarization and helicity. Since the mask operates based on polarization splitting and focusing instead of polarization filtering, it overcomes both the 50$\%$ theoretical efficiency limit, and the one-pixel registration error (resulting from distinct physical areas of the polarization filters) of the previously demonstrated DoFP-PCs~\cite{Tyo2006ApplOpt}. In addition, unlike the previously demonstrated full Stokes DoFP-PCs, the metasurface is fabricated in a single dielectric layer and does not require integration of multiple layers operating as retarders and polarization filters. The mask is designed for 850-nm center wavelength. The polarization cross-talk ranges from 10$\%$ to 15$\%$ for pixel sizes from 7.2~$\mathrm{\mu}$m to 2.4~$\mathrm{\mu}$m when using an 850-nm LED as the light source. In addition, we use a polarization mask to demonstrate that the metasurface DoFP-PC can be used to form polarization images over extended scenes. To the best of our knowledge, this is the first demonstration of a DoFP-PC mask that measures the polarization state completely and is not based on polarization filtering.

\begin{figure*}[htp]

\includegraphics{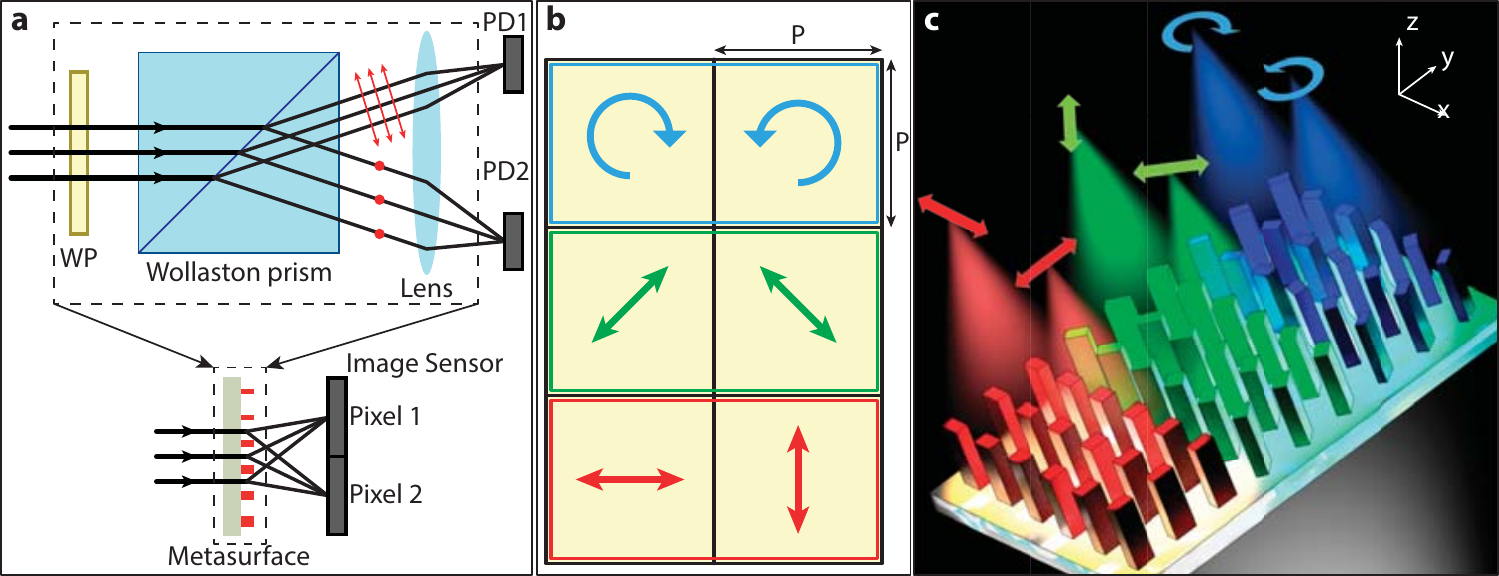}
\caption{Concept of a metasurface polarization camera. \textbf{(a)} Top: Schematics of a conventional setup used for polarimetry: a waveplate (quarter or half) followed by a Wollaston prism and a lens that focuses light on detectors. Bottom: A compact metasurface implements the functionality of all three components combined, and can be directly integrated on an image sensor. WP: waveplate; PD: photodetector. \textbf{(b)} A possible arrangement for a superpixel of the polarization camera, comprising six image sensor pixels. Three independent polarization basis (H/V, $\pm$45$^\circ$, and RHCP/LHCP) are chosen to measure the Stokes parameters at each superpixel. \textbf{(c)} Three-dimensional illustration of a superpixel focusing different polarizations to different spots. The colors are used only for clarity of the image and bear no wavelength information.}
\label{Fig1}
\end{figure*}

There are several representations for polarization of light~\cite{Huard1997Polarization}. Among them, the Stokes vector formalism has some conceptual and experimental advantages since it can be used to represent light with various degrees of polarization, and can be directly determined by measuring the power in certain polarization bases~\cite{Huard1997Polarization}. Therefore, most imaging polarimetry systems determine the Stokes vector~\cite{Tyo2006ApplOpt}, which is usually defined as $S = (1/I) [I,~(I_\mathrm{x}-I_\mathrm{y}),~(I_\mathrm{45}-I_\mathrm{-45}),~(I_\mathrm{R}-I_\mathrm{L})]$, where $I$ is the total intensity, $I_\mathrm{x}$, $I_\mathrm{y}$, $I_\mathrm{45}$, and $I_\mathrm{-45}$ are the intensity of light in linear polarization bases along the x, y, +45-degree, and -45-degree directions, respectively. $I_\mathrm{R}$ and $I_\mathrm{L}$ denote the intensities of the right-hand and left-hand circularly polarized light. To fully characterize the state of polarization, all these intensities should be determined. A conventional setup used to measure the full Stokes vector is shown in \ref{Fig1}a: a waveplate (half or quarter), followed by a Wollaston prism and a lens that focuses the beams on photodetectors. One can determine the four Stokes parameters~\cite{Huard1997Polarization} from the detector signals without a waveplate, with a half-waveplate (HWP), and a quarter-waveplate (QWP). An optical metasurface with the ability to fully control phase and polarization of light~\cite{Arbabi2015NatNano} can perform the same task over a much smaller volume and without changing any optical components. The metasurface can split any two orthogonal states of polarization and simultaneously focus them to different points with high efficiency and on a micron-scale. This is schematically shown in \ref{Fig1}a. Such a metasurface can be directly integrated on an image sensor for making a polarization camera. To fully measure the Stokes parameters, the projection of the input light on three different polarization basis sets should be measured. A typical choice of basis is horizontal/vertical (H/V), $\pm$45$^\circ$ linear, and right-hand-circular/left-hand-circular (RHCP/LHCP) that can be used to directly measure the Stokes parameters. \ref{Fig1}b shows a possible configuration where the three metasurface polarization beam-splitters (PBS) are multiplexed to make a superpixel, comprising of six image sensor pixels. Each image sensor pixel can then be used to measure the power in a single polarization state. A schematic illustration of a superpixel is shown in \ref{Fig1}c. The colors are only used to distinguish different parts of the super pixel more easily, and do not correspond to actual wavelengths. The blue nano-posts, separate and focus RHCP/LHCP, the green ones and the red ones do the same for $\pm$45$^\circ$ and H/V, respectively.

\begin{figure*}[htp]
\centering
\includegraphics[width=1\columnwidth]{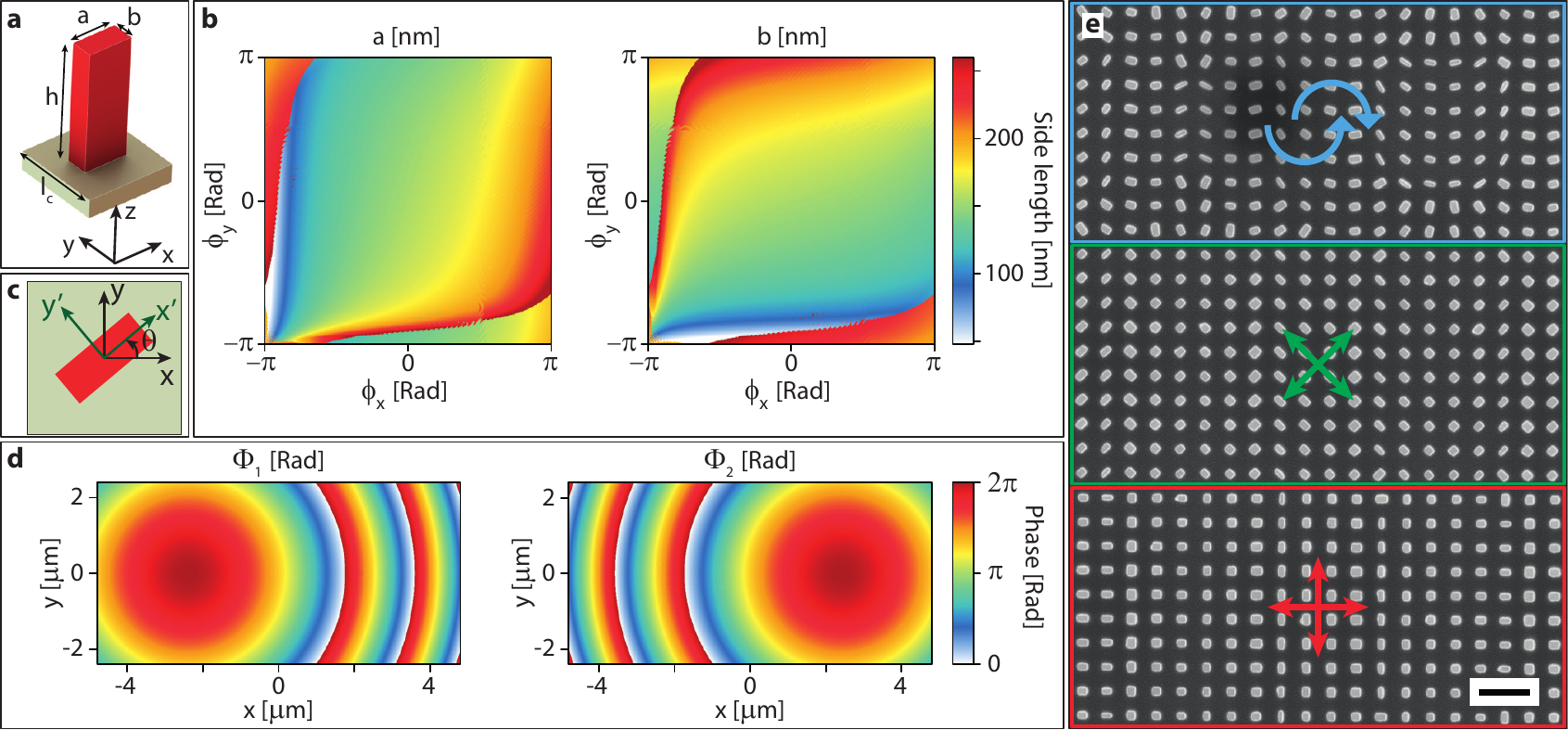}
\caption{Meta-atom and pixel design. \textbf{(a)} An $\alpha$-Si nano-post with a rectangular cross section resting on a glass substrate provides full polarization and phase control. \textbf{(b)} Design graphs used for finding the in-plane dimensions of a nano-post that provides a required pair of transmission phases for the x and y-polarized light. The nano-posts are 650~nm tall, and the lattice constant is 480~nm. \textbf{(c)} Schematic illustration of a rotated nano-post, showing the rotation angle and the old and the new optical axis sets. \textbf{(d)} Required phase profiles for a metasurface that does both polarization beam splitting and focusing at two orthogonal polarizations. These can be any set of orthogonal polarizations, linear or elliptical. The focal distance for these phase profiles is 9.6~$\mathrm{\mu}$m, equal to the width of the superpixel in the x direction. The lateral positions of the focal spots are x$=\pm$2.4~$\mathrm{\mu}$m and y$=$0. \textbf{(e)} Scanning electron micrograph of a fabricated superpixel. The polarization basis for each part is shown with the colored arrows. Scale bar: 1~$\mathrm{\mu}$m.}
\label{Fig2}
\end{figure*}

The metasurface platform we use here is based on the dielectric birefringent nano-post structure~\cite{Arbabi2015NatNano}. As seen in \ref{Fig2}a, the metasurface is composed of $\alpha$-Si nano-posts with rectangular cross-sections on a low-index fused silica substrate. With a proper choice of the $\alpha$-Si layer thickness and lattice constant (650~nm and 480~nm, respectively at a wavelength of 850~nm), the nano-posts can provide full and independent 2$\pi$ phase control over x and y-polarized light where x and y are aligned with the axes of the nano-post (see Supplementary Information Fig. S1)~\cite{Liu2012CompPhys}. Using the phase versus dimension graphs, one could calculate the nano-post dimensions required to provide a specific pair of phase values, $\mathrm{\phi_x}$ and $\mathrm{\phi_y}$, as shown in \ref{Fig2}b. This allows for designing a metasurface that controls x and y-polarized light independently. With a simple generalization, the same can be applied to any two orthogonal linear polarizations using nano-posts that are rotated around their optical axis with the correct angle to match the new linear polarizations (e.g., the x$^\prime$-y$^\prime$ axis in \ref{Fig2}c). An important and interesting point demonstrated in~\cite{Arbabi2015NatNano} is that this can be done on a point-by-point manner, where the polarization basis is different for each nano-post. This property allows us to easily design the metasurface PBS for the two linear bases of interest. Moreover, as demonstrated in~\cite{Arbabi2015NatNano}, an even more interesting property of this seemingly simple structure is that the independent control of orthogonal polarizations can be generalized to elliptical and circular polarizations as well (with a small drawback that will be discussed later). To see this, here we reiterate the results presented in the supplementary material of~\cite{Arbabi2015NatNano}, as it is important to make the design process clear. The operation of a nano-post can be modeled by a Jones matrix relating the input and output electric fields (i.e., $\mathrm{\textbf{E}^{out}}=\mathrm{\textbf{T}}\mathrm{\textbf{E}^{in}}$). For the rotated nano-post shown in \ref{Fig2}c, the Jones matrix can be written as:
\begin{equation}
\mathrm{\textbf{T}} = \left[ {\begin{array}{cc}
   T_{xx} & T_{xy}\\
   T_{yx} & T_{yy}\\
  \end{array} } \right]
=\mathrm{\textbf{R}}(\theta)
  \left[ {\begin{array}{cc}
   \mathrm{e}^{i\phi_{x^\prime}} & 0\\
   0 & \mathrm{e}^{i\phi_{y^\prime}}\\
  \end{array} } \right]
\mathrm{\textbf{R}}(-\theta),
\label{eq:Jones_Matrix}\\
\end{equation}
where $\mathrm{\textbf{R}}(\theta)$ denotes the rotation matrix by an angle $\theta$ in the counter-clockwise direction. Here we have assumed a unity transmission since the nano-posts are highly transmissive. We note here that the right hand side of Equation~\ref{eq:Jones_Matrix} is a unitary and symmetric matrix. Using only these two conditions (i.e., unitarity and symmetry), we find $T_{xy}=T_{yx}$, $|T_{yx}|=\sqrt{1-|T_{xx}|^2}$, and $T_{yy}=-\mathrm{exp(i 2}\angle T_{yx}\mathrm{)}T_{xx}$. As one could expect, these reduce the available number of parameters of the matrix to three ($|T_{xx}|$, $\angle T_{xx}$, $\angle T_{yx}$), corresponding to the three available physical parameters ($\phi_{x^{\prime}}$, $\phi_{y^{\prime}}$, and $\theta$). Using these relations to simplify $\mathrm{\textbf{E}^{out}}=\mathrm{\textbf{T}}\mathrm{\textbf{E}^{in}}$, we can rewrite it to find the Jones matrix elements in terms of the input and output fields:
\begin{equation}
\left[ {\begin{array}{cc}
   E_{x}^\mathrm{out*} & E_{y}^\mathrm{out*}\\
   E_{x}^\mathrm{in} & E_{y}^\mathrm{in}\\
  \end{array} } \right]
\left[ {\begin{array}{c}
   T_{xx}\\
   T_{yx}\\
  \end{array} } \right]
=\left[ {\begin{array}{c}
   E_{x}^\mathrm{in*}\\
   E_{x}^\mathrm{out}\\
  \end{array} } \right],
\label{eq:ElementFinding}
\end{equation}
where $*$ denotes complex conjugation. Equation~\ref{eq:ElementFinding} is important as it shows how one can find the Jones matrix required to transform any input field with a given phase and polarization, to any desired output field with a different phase and polarization. This is the first application of the birefringent meta-atoms, i.e., \textit{complete and independent polarization and phase control}. The Jones matrix is uniquely determined by Equation~\ref{eq:ElementFinding}, unless the determinant of the coefficients matrix on the left hand side of Equation~\ref{eq:ElementFinding} is zero (i.e., the matrix rows are proportional). Since the Jones matrix is unitary (i.e., the input and output powers are equal), the proportionality coefficient must have a unit amplitude: $\mathrm{\textbf{E}_1^{out}}=\mathrm{exp(i}\phi_1\mathrm{)}\mathrm{\textbf{E}_1^{in*}}$, where we are using numeral subscripts to distinguish between two different sets of input/output fields. This means that $\mathrm{\textbf{E}_1^{out}}$ and $\mathrm{\textbf{E}_1^{in}}$ have the same polarization ellipse, but an opposite handedness. In this case, a second equation is required to uniquely determine the Jones matrix. To impose only one equation, the second set of input and output polarizations should also satisfy the same condition as the first ones: $\mathrm{\textbf{E}_2^{out}}=\mathrm{exp(i}\phi_2\mathrm{)}\mathrm{\textbf{E}_2^{in*}}$. If $\phi_1$ and $\phi_2$ can be independently controlled, one can see using a conservation of energy argument, that $\mathrm{\textbf{E}_1^{in}}$ and $\mathrm{\textbf{E}_2^{in}}$ (as well as $\mathrm{\textbf{E}_1^{out}}$ and $\mathrm{\textbf{E}_2^{out}}$) should be orthogonal to each other. Using these two conditions, we can write the final equation as:
\begin{equation}
\left[ {\begin{array}{cc}
   E_{1,x}^\mathrm{in} & E_{1,y}^\mathrm{in}\\
   E_{2,x}^\mathrm{in} & E_{2,y}^\mathrm{in}\\
  \end{array} } \right]
\left[ {\begin{array}{c}
   T_{xx}\\
   T_{yx}\\
  \end{array} } \right]
=\left[ {\begin{array}{c}
   E_{1,x}^\mathrm{out}\\
   E_{2,x}^\mathrm{out}\\
  \end{array} } \right]
=\left[ {\begin{array}{c}
   \mathrm{exp(i}\phi_1\mathrm{)}E_{1,x}^\mathrm{in*}\\
   \mathrm{exp(i}\phi_2\mathrm{)}E_{2,x}^\mathrm{in*}\\
  \end{array} } \right].
\label{eq:IndPhaseCont}
\end{equation}
This is the second important application of the method, \textit{polarization controlled phase manipulation}: given \textit{any} two orthogonal input polarizations (denoted by $\mathrm{\textbf{E}_1^{in}}$ and $\mathrm{\textbf{E}_2^{in}}$), their phase can be independently controlled using the Jones matrix given by Equation~\ref{eq:IndPhaseCont}. For instance, Arbabi et. al.~\cite{Arbabi2015NatNano}, demonstrated a metasurface that focuses RHCP input light to a tight spot, and LHCP input light to a doughnut shape. The cost is that the output orthogonal polarizations have the opposite handedness compared to the input ones. When the Jones matrix is calculated from Equation~\ref{eq:IndPhaseCont} (or Equation~\ref{eq:ElementFinding}, depending on the function), the two phases ($\phi_{x^{\prime}}$ and $\phi_{y^{\prime}}$) and the rotation angle ($\theta$) can be calculated from Equation~\ref{eq:Jones_Matrix}, by finding the eigenvalues and eigenvectors of the Jones matrix. Let us emphasize here that since this is a point-by-point design, all the steps can be repeated independently for each nano-post, meaning that the polarization basis can be changed from one nano-post to the next.

Based on the concept and technique just described, the first design step is identifying the input polarizations at each point. For the DoFP-PC, three different sets of H/V, $\pm$45$^\circ$, and RHCP/LHCP (corresponding to the three distinct areas in the superpixel shown in \ref{Fig1}b) are chosen. Then, the required phase profiles are determined to split each two orthogonal polarizations and focus them to the centers of adjacent pixels on the image sensors (as shown schematically in \ref{Fig1}c). For a pixel size of 4.8~$\mathrm{\mu}$m, the calculated phase profiles are shown in \ref{Fig2}d, where the distance between the metasurface mask and the image sensor is assumed to be 9.6~$\mathrm{\mu}$m. Since each polarization basis covers two image sensor pixels, the phases are defined over the area of two pixels. In addition, the calculated phases are the same for the three different polarization bases, and therefore only one basis set is shown in \ref{Fig2}d. Using these phases and knowing the desired polarization basis at each point, we calculated the rotation angles and nano-post dimensions from Equations~\ref{eq:IndPhaseCont} and \ref{eq:Jones_Matrix} along with the data shown in~\ref{Fig2}b.

The metasurface mask was then fabricated in a process similar to Ref.~\cite{Arbabi2016OptExp}. A 650-nm-thick layer of $\alpha$-Si was deposited on a fused silica wafer. The metasurface pattern was defined using electron-beam lithography, and transferred to the $\alpha$-Si layer through a lift-off process (to make a hard etch-mask) followed by dry etching. \ref{Fig2}e shows a scanning electron micrograph of a fabricated superpixel, with the polarization bases denoted by arrows for each section. In addition to the metasurface mask corresponding to a pixel size of 4.8~$\mathrm{\mu}$m (mentioned above and shown in \ref{Fig2}e), two other masks with pixel sizes of 7.2~$\mathrm{\mu}$m and 2.4~$\mathrm{\mu}$m were also fabricated (with metasurface to image sensor separations of 14.4~$\mathrm{\mu}$m and 4.8~$\mathrm{\mu}$m, respectively) to study the effect of pixel size on the imaging performance.

\begin{figure*}[htp]
\centering
\includegraphics[width=1\columnwidth]{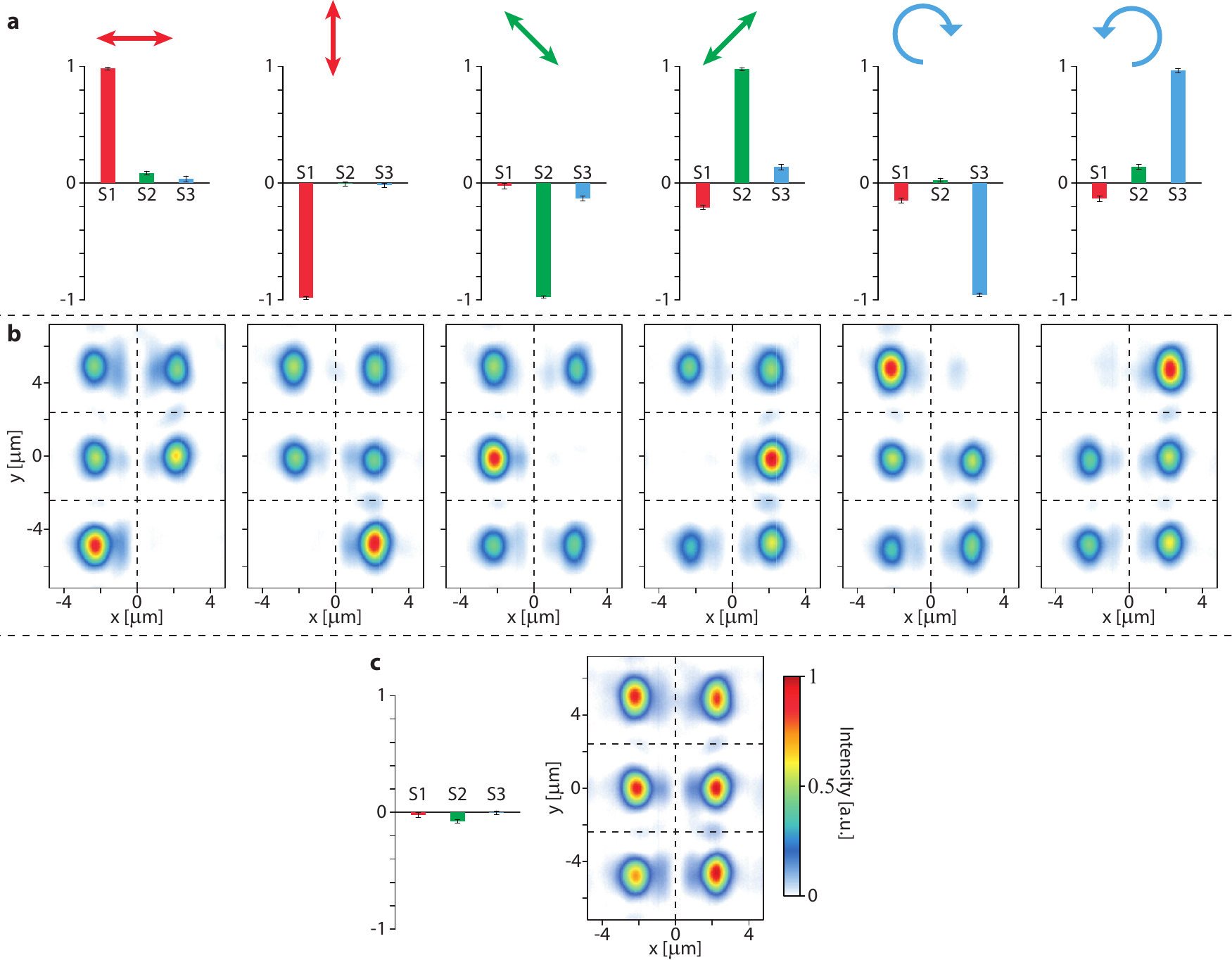}
\caption{Characterization results of the superpixels of the DoFP metasurface mask. \textbf{(a)} Calculated average Stokes parameters for different input polarizations (shown with colored arrows), and \textbf{(b)} the corresponding intensity distributions for a sample superpixel. The Stokes parameters are averaged over about 120 superpixels (limited by the microscope field of view), and the error bars represent the statistical standard deviations. \textbf{(c)} Calculated Stokes parameters and the corresponding intensity distribution for the LED light source without any polarization filters in the setup. All the measurements are performed with an 850-nm LED filtered by a bandpass filter (center: 850~nm, FHMW: 10~nm) as the light source.}
\label{Fig3}
\end{figure*}

To characterize the metasurface mask, we illuminated it with light from an 850-nm LED (filtered by a 10-nm bandpass filter) with different states of polarization, and imaged the plane corresponding to the image sensor location using a custom-built microscope (see Supplementary Fig. S2 for measurement details and the setup). \ref{Fig3} summarizes the superpixel characterization results for the 4.8-$\mathrm{\mu}$m pixel design. The measured Stokes parameters are plotted in~\ref{Fig3}a for different input polarizations showing results with low cross-talk (<10$\%$) between polarizations and high similarity between different superpixels. The measurements were averaged over more than 120 superpixels (limited by the field of view of the microscope), and the standard deviations are shown in the graph as error bars. In addition, the intensity distribution over a sample superpixel area is shown in \ref{Fig3}b for the same input polarizations. The graphs show the clear ability of the metasurface mask to route light as desired for various input polarizations. Similar characterization results without a bandpass filter (i.e., for a bandwith of about 5$\%$) are presented in Supplementary Information Figure S3, showing slight performance degradation (with a maximum cross-talk of $\sim$13$\%$) as the metasurface efficiency decreases with changing the wavelength. In addition, similar measurement results for metasurface masks with pixel sizes of 7.2~$\mathrm{\mu}$m and 2.4~$\mathrm{\mu}$m are presented in Supplementary Information Figures S4 and S5, respectively. The results show a degradation of performance with reducing the pixel size (the cross-talk is smaller than 7.5$\%$ and 13$\%$ for 7.2-$\mathrm{\mu}$m and 2.4-$\mathrm{\mu}$m pixels, respectively). To show the ability of the metasurface mask to characterize the polarization state of unpolarized light, we repeated the same measurements with the polarization filter removed from the setup. \ref{Fig3}c summarizes the results of this measurement that determines the state of polarization of light emitted by the LED. The data given in \ref{Fig3}a is used to estimate the calibration matrix. As expected, the emitted light has a low degree of polarization ($<$0.08). We also characterized the polarization state of the emitted LED light using a QWP and an LP, and found the degree of polarization to be equal to zero upto the measurement error.

In addition, we measured the transmission efficiency of the metasurface mask and found it to be in the range of 60$\%$ to 65$\%$ for all pixel size designs and input polarizations. The lower than expected transmission is mainly due to a few factors. First, the metasurface has a maximum deflection angle larger than 50$^\circ$, which results in lower transmission efficiency~\cite{Arbabi2015NatCommun,Arbabi2017SPIEPW}. Second, the relatively large metasurface lattice constant of 480~nm does not satisfy the Nyquist sampling theorem for the large-deflection-angle transmission masks inside the fused silica substrate~\cite{Kamali2016LaserPhotonRev}. This results in spurious diffraction of light inside the substrate. Finally, the mask is periodic with a larger-than-wavelength period equal to the superpixel dimensions. This results in excitation of higher diffraction orders especially inside the substrate that has a higher refractive index. It is worth noting that the achieved $\sim$65$\%$ efficiency is higher than the theoretical limit of a polarimetric camera that is based on polarization filtering (e.g., uses a nano-wire grid polarizer).

\begin{figure*}[htp]
\centering
\includegraphics[width=1\columnwidth]{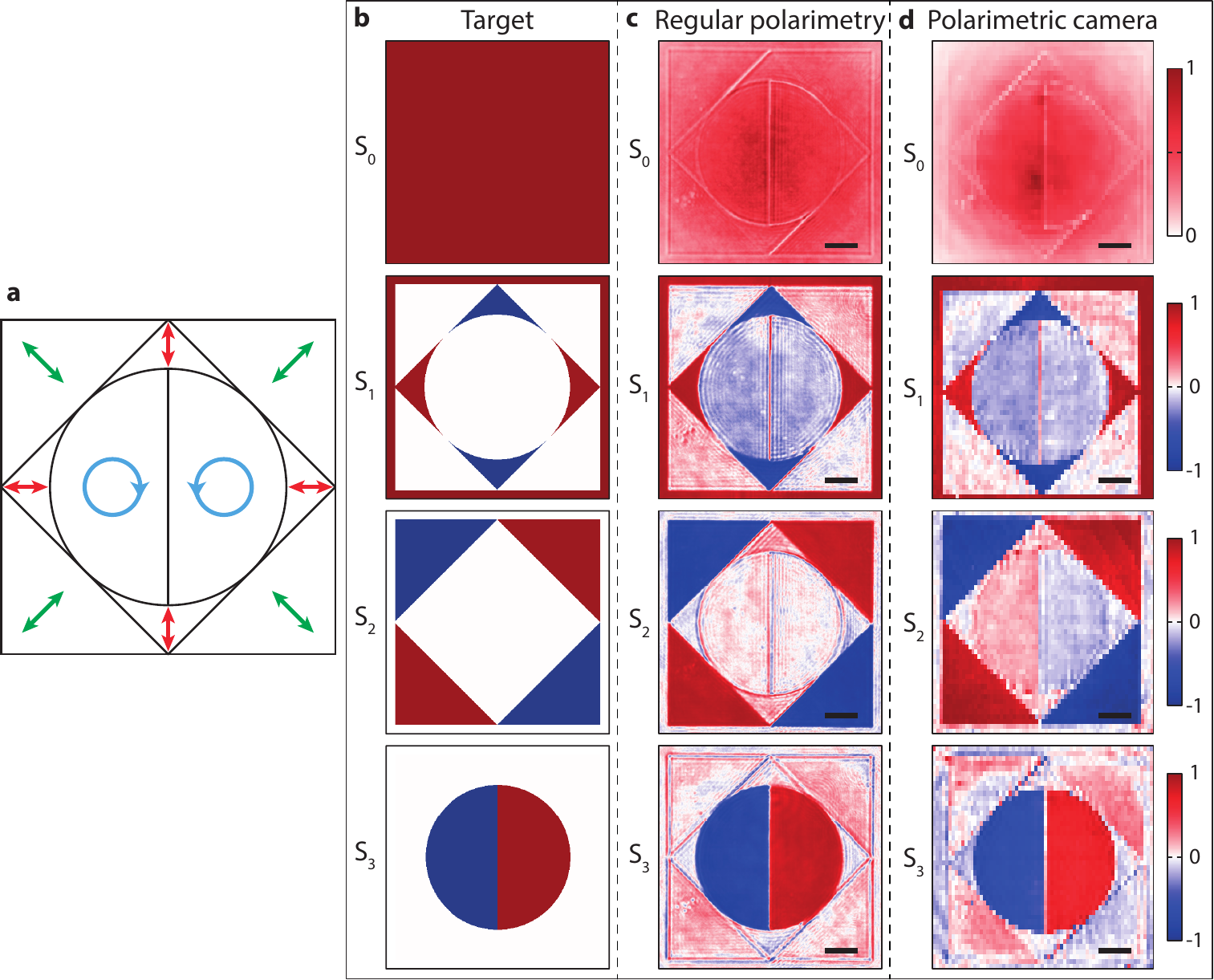}
\caption{Polarimetric imaging. \textbf{(a)} Schematic illustration of target polarization ellipse in different parts of the polarization sample. Stokes parameters of the polarization sample: \textbf{(b)} the targeted polarization mask, \textbf{(c)} the fabricated mask imaged using conventional polarimetry, and \textbf{(d)} the same mask imaged using the metasurface polarimetric camera. The scale bars denote 100~$\mathrm{\mu}$m in the metasurface polarization camera mask plane.}
\label{Fig4}
\end{figure*}

Finally, we show that using the DoFP metasurface mask, one could perform polarimetric imaging. To do this, we designed and fabricated a metasurface polarization mask (using the polarization-phase control method described above, and a fabrication process similar to the DoFP metasurface mask). The mask converts x-polarized input light to an output polarization state characterized by the polarization ellipses and the Stokes parameters shown in \ref{Fig4}a and \ref{Fig4}b, respectively. Each Stokes parameter is +1 or -1 in an area of the image corresponding to the specific polarization (e.g., S$_3$ being +1 in the right half circle and -1 in the left half circle and 0 elsewhere). Using a second custom-built microscope, the image of the polarization mask was projected on the DoFP metasurface mask (see Supplementary Information Figure S2 for the measurement setup and the details). First, we removed the metasurface mask and performed a conventional polarimetric imaging of the projected image using a linear polarizer (LP) and a QWP. The results are shown in \ref{Fig4}c. Second, we removed the LP and QWP, and inserted the DoFP metasurface mask in its place. The Stokes parameters were extracted from a single image captured at the location of the image-sensor plane in front of the DoFP metasurface mask. The results are shown in \ref{Fig4}d, and are in good agreement with the results of regular polarimetric imaging. The lower quality of the metasurface polarimetric camera image is mainly due to the limited number of superpixels that fit inside the single field of view of the microscope (limited by the microscope magnification and image sensor size, $\times$22 and $\sim$15~mm, respectively). This results in a low resolution of 70-by-46 points for the metasurface polarimetric image versus the $\sim$2000-by-2000 point resolution of the regular polarimetric image. In addition, to form the  final image, we need to know the coordinates of each superpixel a priori. The existing errors in estimating these coordinates (resulting from small tilts in the setup, aberrations of the custom-built microscope, etc.) cause a degraded performance over some superpixels. In a polarization camera made using the metasurface DoFP metasurface mask, both of these issues will be solved as the resolution can be much higher, and the mask and the image sensor are lithographically aligned.

To extract the polarization information of the image, we integrated the intensity inside the area of two adjacent image sensor pixels, and calculated the corresponding Stokes parameter simply by dividing their difference by their sum. While straightforward, this is not the optimal method to perform this task as there is non-negligible cross-talk between different polarization intensities measured by the pixels (\ref{Fig3}). The issue becomes more important as one moves toward smaller pixel sizes (e.g., the 2.4-$\mathrm{\mu}$m pixel of Supplementary Information Figure S5). To address this, a better polarization data extraction method is to form a matrix that relates the actual intensity of different input polarizations to the corresponding measured values for a specific DoFP metasurface mask design (for instance using the data in \ref{Fig3}). This allows one to reduce the effect of the cross-talk and measure the polarization state more precisely.

The designed small distance between the metasurface and the image sensor (e.g., 9.6~$\mathrm{\mu}$m for the 4.8-$\mathrm{\mu}$m pixel) results in a diffraction-limited bandwidth of about 40$\%$ (assuming a constant phase profile that doesn't change with wavelength and using the criterion given in~\cite{Arbabi2016NatCommun}). Therefore, the actual bandwidth of the device is limited by the focusing and polarization control efficiencies that drop with detuning from the design wavelength. In addition, it is expected that the same level of performance achieved from the 2.4-$\mathrm{\mu}$m pixel in this work, can be achieved from a $\sim$1.7-$\mathrm{\mu}$m pixel if the material between the mask and the image sensor has a refractive index of 1.5, which is the case when the DoFP mask is separated from the image sensor by an oxide or polymer layer, as in a realistic device. To achieve smaller pixel sizes, better performance, and larger operation bandwidths one could use more advanced optimization~\cite{Sell2017NanoLett} or chromatic-dispersion control techniques~\cite{Arbabi2017Optica}, especially since the size of a single superpixel is small and allows for a fast simulation of the forward problem. In addition, a spatial multiplexing scheme~\cite{Bayer1976Patent,Arbabi2016SciRep,Lin2016NanoLett} can be used to interleave multiple superpixels corresponding to different optical bands, and therefore make a color-polarization camera.

Using the polarization-phase control method and the platform introduced in~\cite{Arbabi2015NatNano}, we demonstrated a metasurface mask for DoFP-PCs. The mask is designed to split and focus light to six different pixels on an image sensor for three different polarization bases. This allows for complete characterization of polarization by measuring the four Stokes parameters over the area of each superpixel, which corresponds to the area of six pixels on the image sensor. We experimentally demonstrated the ability of the metasurface masks to correctly measure the state of polarization for different input polarizations. In addition, we used the DoFP metasurface mask to form an image of a complicated polarization object, showing the ability to make a polarization camera. Many of the limitations faced here can be overcome using more advanced optimization techniques or better data extraction methods. We anticipate that polarization cameras based on metasurface masks will be able to replace the conventional polarization cameras for many applications as they enable measurement of the full polarization state including the degree of circular polarization and handedness.

\section*{Materials and Methods}
\noindent\textbf{Simulation and design.}
To design the DoFP metasurface mask, we first calculated the two phase profiles required for the two polarization states [\ref{Fig2}d]. The phase profiles correspond to decentered aspheric lenses that focus each polarization at the center of one image sensor pixel. These phases are then used in Equation~\ref{eq:IndPhaseCont} along with the known input polarization states to calculate the Jones matrix. To find the nano-post corresponding to each Jones matrix, the matrix is diagonalized according to Equation~\ref{eq:Jones_Matrix}, and the two phases ($\phi_{x^{\prime}}$ and $\phi_{y^{\prime}}$) and the rotation angle $\theta$ are then extracted. The nano-post providing the required pair of phases is then found using the data in \ref{Fig2}b.

The polarization target used for the imaging experiments in \ref{Fig4} was designed in a slightly different manner, since in this case only the output polarization is of interest. Assuming an \textit{x}-polarized input light, the output polarization at each point on the mask was chosen according to \ref{Fig4}a. In the general case, the mask can then be designed using the Jones matrix found from Equation~\ref{eq:ElementFinding}, and calculating the corresponding phases and rotation from the Jones matrix. In this especial case, however, the device is a set of nano-posts acting as quarter or half wave-plates. Therefore, we deigned the nano-posts in a manner similar to~\cite{Backlund2016NatPhoton} to make it robust to fabrication errors.

To find the transmission amplitude and phase for the nano-posts [Supplementary Fig. S1], we simulated a uniform array of nano-posts with rectangular cross-sections under normally incident \textit{x}- and \textit{y}-polarized light using the rigorous coupled wave analysis~\cite{Liu2012CompPhys}. The resulting complex transmissions were then used to find the best nano-post that provides each required phase pair through minimizing the Euclidean distance between $[e^{i\phi_x}, e^{i\phi_y}]$ and $[t_x, t_y]$, where $\phi_x$ and $\phi_y$ are the desired phase values, and $t_x$ and $t_y$ are complex nano-post transmissions. The optimized nano-post dimensions are plotted in \ref{Fig2}b.

\noindent\textbf{Fabrication.}
The fabrication process is the same for both the DoFP metasurface mask and the polarization imaging target. The fabrication started with deposition of a 650-nm-thick layer of $\alpha$-Si on a 500-$\mu$m-thick fused silica substrate. The metasurface pattern is defined in a $\sim$300-nm-thick ZEP-520A positive electron-beam resist using electron-beam lithography. After development of the resist, a $\sim$70-nm-thick layer of aluminum oxide is deposited on the sample using electron-beam evaporation and lifted off to invert the pattern. The aluminum oxide is then used as a hard mask in the reactive ion etching of the $\alpha$-Si layer. Finally, the aluminum oxide mask is removed in a solution of hydrogen peroxide and ammonium hydroxide.

\noindent\textbf{Measurement.}
The measurement setups (including part models) are schematically illustrated in Supplementary Fig. S2 for different parts of the characterization process. To characterize the DoFP super-pixel performance, light from an LED was passed through an LP and a QWP to set the input polarization state. The six different polarization states [\ref{Fig3}a] were generated using this combination. The intensity distribution patterns at the focal plane after the DoFP metasurface mask were then imaged using a custom-built microscope. The data was analyzed by calculating the Stokes parameters measured by each super-pixel, and averaging over all the super-pixels that fit within the field of view. A 10-nm bandwidth filter with a center wavelength of 850~nm was inserted in the path to characterize the narrow-band operation, and was then removed to acquire the results for a wider-bandwidth source.

The imaging polarimetry experiments were performed in a similar way. For these experiments, the polarization target was illuminated by \textit{x}-polarized light out of a supercontinuum laser source (filtered by the same 10-nm bandwidth filter). The target was imaged onto the DoFP metasurface mask plane using a secondary custom-built microscope (operating as relay optics). The intensity distribution at the focal plane after the DoFP metasurface mask was then imaged and analyzed to generate the polarization images plotted in \ref{Fig4}d. For comparison, the DoFP metasurface mask was removed and a polarization analyzer (i.e., a QWP and an LP) was inserted into the system to form the reference polarization images plotted in \ref{Fig4}c.

\section{References}
\bibliographystyle{naturemag_noURL}
\bibliography{MetasurfaceLibrary}

\begin{thebibliography}{10}
\expandafter\ifx\csname url\endcsname\relax
  \def\url#1{\texttt{#1}}\fi
\expandafter\ifx\csname urlprefix\endcsname\relax\def\urlprefix{URL }\fi
\providecommand{\bibinfo}[2]{#2}
\providecommand{\eprint}[2][]{\url{#2}}

\bibitem{Chang2003Hyperspectral}
\bibinfo{author}{Chang, C.-I.}
\newblock \emph{\bibinfo{title}{Hyperspectral imaging: techniques for spectral
  detection and classification}}, vol.~\bibinfo{volume}{1}
  (\bibinfo{publisher}{Springer Science $\&$ Business Media},
  \bibinfo{year}{2003}).

\bibitem{Sun2010Hyperspectral}
\bibinfo{author}{Sun, D.-W.}
\newblock \emph{\bibinfo{title}{Hyperspectral imaging for food quality analysis
  and control}} (\bibinfo{publisher}{Elsevier}, \bibinfo{year}{2010}).

\bibitem{Tyo2006ApplOpt}
\bibinfo{author}{Tyo, J.~S.}, \bibinfo{author}{Goldstein, D.~L.},
  \bibinfo{author}{Chenault, D.~B.} \& \bibinfo{author}{Shaw, J.~A.}
\newblock \bibinfo{title}{Review of passive imaging polarimetry for remote
  sensing applications}.
\newblock \emph{\bibinfo{journal}{Appl. Opt.}} \textbf{\bibinfo{volume}{45}},
  \bibinfo{pages}{5453--5469} (\bibinfo{year}{2006}).

\bibitem{Garcia2015OptExpress}
\bibinfo{author}{Garcia, N.~M.}, \bibinfo{author}{de~Erausquin, I.},
  \bibinfo{author}{Edmiston, C.} \& \bibinfo{author}{Gruev, V.}
\newblock \bibinfo{title}{Surface normal reconstruction using circularly
  polarized light}.
\newblock \emph{\bibinfo{journal}{Opt. Express}} \textbf{\bibinfo{volume}{23}},
  \bibinfo{pages}{14391--14406} (\bibinfo{year}{2015}).

\bibitem{Coffeen1979JOSA}
\bibinfo{author}{Coffeen, D.~L.}
\newblock \bibinfo{title}{Polarization and scattering characteristics in the
  atmospheres of earth, venus, and jupiter}.
\newblock \emph{\bibinfo{journal}{J. Opt. Soc. Am.}}
  \textbf{\bibinfo{volume}{69}}, \bibinfo{pages}{1051--1064}
  (\bibinfo{year}{1979}).

\bibitem{Walraven1981SPIEOptEng}
\bibinfo{author}{Walraven, R.}
\newblock \bibinfo{title}{Polarization imagery}.
\newblock In \emph{\bibinfo{booktitle}{SPIE Opt. Eng.}},
  vol.~\bibinfo{volume}{20}, \bibinfo{pages}{5} (\bibinfo{publisher}{SPIE},
  \bibinfo{year}{1981}).

\bibitem{Egan1991ApplOpt}
\bibinfo{author}{Egan, W.~G.}, \bibinfo{author}{Johnson, W.~R.} \&
  \bibinfo{author}{Whitehead, V.~S.}
\newblock \bibinfo{title}{Terrestrial polarization imagery obtained from the
  space shuttle:characterization and interpretation}.
\newblock \emph{\bibinfo{journal}{Appl. Opt.}} \textbf{\bibinfo{volume}{30}},
  \bibinfo{pages}{435--442} (\bibinfo{year}{1991}).

\bibitem{Liu2012JBioOpt}
\bibinfo{author}{Liu, Y.} \emph{et~al.}
\newblock \bibinfo{title}{Complementary fluorescence-polarization microscopy
  using division-of-focal-plane polarization imaging sensor}.
\newblock \emph{\bibinfo{journal}{J. Biomed. Opt.}}
  \textbf{\bibinfo{volume}{17}}, \bibinfo{pages}{116001}
  (\bibinfo{year}{2012}).

\bibitem{Roberts2014ProcIEEE}
\bibinfo{author}{Roberts, N.~W.} \emph{et~al.}
\newblock \bibinfo{title}{Animal polarization imaging and implications for
  optical processing}.
\newblock \emph{\bibinfo{journal}{Proceed. IEEE}}
  \textbf{\bibinfo{volume}{102}}, \bibinfo{pages}{1427--1434}
  (\bibinfo{year}{2014}).

\bibitem{Charanya2014SPIEJBioOpt}
\bibinfo{author}{Charanya, T.} \emph{et~al.}
\newblock \bibinfo{title}{Trimodal color-fluorescence-polarization endoscopy
  aided by a tumor selective molecular probe accurately detects flat lesions in
  colitis-associated cancer}.
\newblock \emph{\bibinfo{journal}{J. Biomed. Opt.}}
  \textbf{\bibinfo{volume}{19}}, \bibinfo{pages}{14} (\bibinfo{year}{2014}).

\bibitem{Garcia2017Optica}
\bibinfo{author}{Garcia, M.}, \bibinfo{author}{Edmiston, C.},
  \bibinfo{author}{Marinov, R.}, \bibinfo{author}{Vail, A.} \&
  \bibinfo{author}{Gruev, V.}
\newblock \bibinfo{title}{Bio-inspired color-polarization imager for real-time
  in situ imaging}.
\newblock \emph{\bibinfo{journal}{Optica}} \textbf{\bibinfo{volume}{4}},
  \bibinfo{pages}{1263--1271} (\bibinfo{year}{2017}).

\bibitem{Johnson1974AMERDC}
\bibinfo{author}{Johnson, J.}
\newblock \bibinfo{title}{Infrared polarization signature feasibility tests}.
\newblock \emph{\bibinfo{journal}{US Army Mobility Equipment Research and
  Development Center}} \textbf{\bibinfo{volume}{TR-EO-74-1 (AD COO1-133)}}
  (\bibinfo{year}{1974}).

\bibitem{Chin-Bing1976DTIC}
\bibinfo{author}{Chin-Bing, S.}
\newblock \bibinfo{title}{Infrared polarization signature analysis}.
\newblock \emph{\bibinfo{journal}{Defense Technical Information Center}}
  \textbf{\bibinfo{volume}{Rep. ADC008418}} (\bibinfo{year}{1976}).

\bibitem{Solomon1981ApplOpt}
\bibinfo{author}{Solomon, J.~E.}
\newblock \bibinfo{title}{Polarization imaging}.
\newblock \emph{\bibinfo{journal}{Appl. Opt.}} \textbf{\bibinfo{volume}{20}},
  \bibinfo{pages}{1537--1544} (\bibinfo{year}{1981}).

\bibitem{Pezzaniti1995OptEng}
\bibinfo{author}{Pezzaniti, J.~L.} \& \bibinfo{author}{Chipman, R.~A.}
\newblock \bibinfo{title}{Mueller matrix imaging polarimetry}.
\newblock In \emph{\bibinfo{booktitle}{Opt. Eng.}}, vol.~\bibinfo{volume}{34},
  \bibinfo{pages}{11} (\bibinfo{publisher}{SPIE}, \bibinfo{year}{1995}).

\bibitem{Nordin1999JOSAA}
\bibinfo{author}{Nordin, G.~P.}, \bibinfo{author}{Meier, J.~T.},
  \bibinfo{author}{Deguzman, P.~C.} \& \bibinfo{author}{Jones, M.~W.}
\newblock \bibinfo{title}{Micropolarizer array for infrared imaging
  polarimetry}.
\newblock \emph{\bibinfo{journal}{J. Opt. Soc. Am. A}}
  \textbf{\bibinfo{volume}{16}}, \bibinfo{pages}{1168--1174}
  (\bibinfo{year}{1999}).

\bibitem{Guo2000ApplOpt}
\bibinfo{author}{Guo, J.} \& \bibinfo{author}{Brady, D.}
\newblock \bibinfo{title}{Fabrication of thin-film micropolarizer arrays for
  visible imaging polarimetry}.
\newblock \emph{\bibinfo{journal}{Appl. Opt.}} \textbf{\bibinfo{volume}{39}},
  \bibinfo{pages}{1486--1492} (\bibinfo{year}{2000}).

\bibitem{Gruev2006IEEESCS}
\bibinfo{author}{Gruev, V.}, \bibinfo{author}{Spiegel, J. V.~d.} \&
  \bibinfo{author}{Engheta, N.}
\newblock \bibinfo{title}{Image sensor with focal plane extraction of
  polarimetric information}.
\newblock In \emph{\bibinfo{booktitle}{2006 IEEE International Symposium on
  Circuits and Systems}}, \bibinfo{pages}{4 pp.--216} (\bibinfo{year}{2006}).

\bibitem{Rust1995Patent}
\bibinfo{author}{Rust, D.}
\newblock \bibinfo{title}{Integrated dual imaging detector}.
\newblock \emph{\bibinfo{journal}{$\mathrm{US Patent}$}}
  \bibinfo{pages}{5,438,414} (\bibinfo{year}{1995}).

\bibitem{Andreou2002IEEESensJour}
\bibinfo{author}{Andreou, A.~G.} \& \bibinfo{author}{Kalayjian, Z.~K.}
\newblock \bibinfo{title}{Polarization imaging: principles and integrated
  polarimeters}.
\newblock \emph{\bibinfo{journal}{IEEE Sens. Jour.}}
  \textbf{\bibinfo{volume}{2}}, \bibinfo{pages}{566--576}
  (\bibinfo{year}{2002}).

\bibitem{Gruev2007OptExp}
\bibinfo{author}{Gruev, V.}, \bibinfo{author}{Ortu, A.},
  \bibinfo{author}{Lazarus, N.}, \bibinfo{author}{Spiegel, J. V.~d.} \&
  \bibinfo{author}{Engheta, N.}
\newblock \bibinfo{title}{Fabrication of a dual-tier thin film
  micropolarization array}.
\newblock \emph{\bibinfo{journal}{Opt. Exp.}} \textbf{\bibinfo{volume}{15}},
  \bibinfo{pages}{4994--5007} (\bibinfo{year}{2007}).

\bibitem{Gruev2010OptExp}
\bibinfo{author}{Gruev, V.}, \bibinfo{author}{Perkins, R.} \&
  \bibinfo{author}{York, T.}
\newblock \bibinfo{title}{Ccd polarization imaging sensor with aluminum
  nanowire optical filters}.
\newblock \emph{\bibinfo{journal}{Opt. Express}} \textbf{\bibinfo{volume}{18}},
  \bibinfo{pages}{19087--19094} (\bibinfo{year}{2010}).

\bibitem{Deguzman2001ApplOpt}
\bibinfo{author}{Deguzman, P.~C.} \& \bibinfo{author}{Nordin, G.~P.}
\newblock \bibinfo{title}{Stacked subwavelength gratings as circular
  polarization filters}.
\newblock \emph{\bibinfo{journal}{Appl. Opt.}} \textbf{\bibinfo{volume}{40}},
  \bibinfo{pages}{5731--5737} (\bibinfo{year}{2001}).

\bibitem{Myhre2012OptExpress}
\bibinfo{author}{Myhre, G.} \emph{et~al.}
\newblock \bibinfo{title}{Liquid crystal polymer full-stokes division of focal
  plane polarimeter}.
\newblock \emph{\bibinfo{journal}{Opt. Express}} \textbf{\bibinfo{volume}{20}},
  \bibinfo{pages}{27393--27409} (\bibinfo{year}{2012}).

\bibitem{Hsu2014OptExpress}
\bibinfo{author}{Hsu, W.-L.} \emph{et~al.}
\newblock \bibinfo{title}{Full-stokes imaging polarimeter using an array of
  elliptical polarizer}.
\newblock \emph{\bibinfo{journal}{Opt. Express}} \textbf{\bibinfo{volume}{22}},
  \bibinfo{pages}{3063--3074} (\bibinfo{year}{2014}).

\bibitem{Hsu2015OptExpress}
\bibinfo{author}{Hsu, W.-L.} \emph{et~al.}
\newblock \bibinfo{title}{Polarization microscope using a near infrared
  full-stokes imaging polarimeter}.
\newblock \emph{\bibinfo{journal}{Opt. Express}} \textbf{\bibinfo{volume}{23}},
  \bibinfo{pages}{4357--4368} (\bibinfo{year}{2015}).

\bibitem{Kuznetsov2016Science}
\bibinfo{author}{Kuznetsov, A.~I.}, \bibinfo{author}{Miroshnichenko, A.~E.},
  \bibinfo{author}{Brongersma, M.~L.}, \bibinfo{author}{Kivshar, Y.~S.} \&
  \bibinfo{author}{Luk’yanchuk, B.}
\newblock \bibinfo{title}{Optically resonant dielectric nanostructures}.
\newblock \emph{\bibinfo{journal}{Science}} \textbf{\bibinfo{volume}{354}}
  (\bibinfo{year}{2016}).

\bibitem{Staude2017NatPhoton}
\bibinfo{author}{Staude, I.} \& \bibinfo{author}{Schilling, J.}
\newblock \bibinfo{title}{Metamaterial-inspired silicon nanophotonics}.
\newblock \emph{\bibinfo{journal}{Nat. Photon.}} \textbf{\bibinfo{volume}{11}},
  \bibinfo{pages}{274--284} (\bibinfo{year}{2017}).

\bibitem{Hsiao2017SmallMeth}
\bibinfo{author}{Hsiao, H.-H.}, \bibinfo{author}{Chu, C.~H.} \&
  \bibinfo{author}{Tsai, D.~P.}
\newblock \bibinfo{title}{Fundamentals and applications of metasurfaces}.
\newblock \emph{\bibinfo{journal}{Small Methods}} \textbf{\bibinfo{volume}{1}},
  \bibinfo{pages}{1600064} (\bibinfo{year}{2017}).

\bibitem{Astilean1998OptLett}
\bibinfo{author}{Astilean, S.}, \bibinfo{author}{Lalanne, P.},
  \bibinfo{author}{Chavel, P.}, \bibinfo{author}{Cambril, E.} \&
  \bibinfo{author}{Launois, H.}
\newblock \bibinfo{title}{High-efficiency subwavelength diffractive element
  patterned in a high-refractive-index material for 633~nm}.
\newblock \emph{\bibinfo{journal}{Opt. Lett.}} \textbf{\bibinfo{volume}{23}},
  \bibinfo{pages}{552--554} (\bibinfo{year}{1998}).

\bibitem{Lalanne1998OptLett}
\bibinfo{author}{Lalanne, P.}, \bibinfo{author}{Astilean, S.},
  \bibinfo{author}{Chavel, P.}, \bibinfo{author}{Cambril, E.} \&
  \bibinfo{author}{Launois, H.}
\newblock \bibinfo{title}{Blazed binary subwavelength gratings with
  efficiencies larger than those of conventional \'{e}chelette gratings}.
\newblock \emph{\bibinfo{journal}{Opt. Lett.}} \textbf{\bibinfo{volume}{23}},
  \bibinfo{pages}{1081--1083} (\bibinfo{year}{1998}).

\bibitem{Fattal2010NatPhoton}
\bibinfo{author}{Fattal, D.}, \bibinfo{author}{Li, J.}, \bibinfo{author}{Peng,
  Z.}, \bibinfo{author}{Fiorentino, M.} \& \bibinfo{author}{Beausoleil, R.~G.}
\newblock \bibinfo{title}{Flat dielectric grating reflectors with focusing
  abilities}.
\newblock \emph{\bibinfo{journal}{Nat. Photon.}} \textbf{\bibinfo{volume}{4}},
  \bibinfo{pages}{466--470} (\bibinfo{year}{2010}).

\bibitem{Lu2010OptExp}
\bibinfo{author}{Lu, F.}, \bibinfo{author}{Sedgwick, F.~G.},
  \bibinfo{author}{Karagodsky, V.}, \bibinfo{author}{Chase, C.} \&
  \bibinfo{author}{Chang-Hasnain, C.~J.}
\newblock \bibinfo{title}{Planar high-numerical-aperture low-loss focusing
  reflectors and lenses using subwavelength high contrast gratings}.
\newblock \emph{\bibinfo{journal}{Opt. Express}} \textbf{\bibinfo{volume}{18}},
  \bibinfo{pages}{12606--12614} (\bibinfo{year}{2010}).

\bibitem{Vo2014IEEEPhotonTechLett}
\bibinfo{author}{Vo, S.} \emph{et~al.}
\newblock \bibinfo{title}{Sub-wavelength grating lenses with a twist}.
\newblock \emph{\bibinfo{journal}{IEEE Photon. Technol. Lett.}}
  \textbf{\bibinfo{volume}{26}}, \bibinfo{pages}{1375--1378}
  (\bibinfo{year}{2014}).

\bibitem{Lin2014Science}
\bibinfo{author}{Lin, D.}, \bibinfo{author}{Fan, P.}, \bibinfo{author}{Hasman,
  E.} \& \bibinfo{author}{Brongersma, M.~L.}
\newblock \bibinfo{title}{Dielectric gradient metasurface optical elements}.
\newblock \emph{\bibinfo{journal}{Science}} \textbf{\bibinfo{volume}{345}},
  \bibinfo{pages}{298--302} (\bibinfo{year}{2014}).

\bibitem{Arbabi2015NatCommun}
\bibinfo{author}{Arbabi, A.}, \bibinfo{author}{Horie, Y.},
  \bibinfo{author}{Ball, A.~J.}, \bibinfo{author}{Bagheri, M.} \&
  \bibinfo{author}{Faraon, A.}
\newblock \bibinfo{title}{Subwavelength-thick lenses with high numerical
  apertures and large efficiency based on high-contrast transmitarrays}.
\newblock \emph{\bibinfo{journal}{Nat. Commun.}} \textbf{\bibinfo{volume}{6}},
  \bibinfo{pages}{7069} (\bibinfo{year}{2015}).

\bibitem{Khorasaninejad2016Science}
\bibinfo{author}{Khorasaninejad, M.} \emph{et~al.}
\newblock \bibinfo{title}{Metalenses at visible wavelengths:
  Diffraction-limited focusing and subwavelength resolution imaging}.
\newblock \emph{\bibinfo{journal}{Science}} \textbf{\bibinfo{volume}{352}},
  \bibinfo{pages}{1190--1194} (\bibinfo{year}{2016}).

\bibitem{Yin2017LighSciAppl}
\bibinfo{author}{Yin, X.} \emph{et~al.}
\newblock \bibinfo{title}{Beam switching and bifocal zoom lensing using active
  plasmonic metasurfaces}.
\newblock \emph{\bibinfo{journal}{Light: Sci. Appl.}}
  \textbf{\bibinfo{volume}{6}}, \bibinfo{pages}{e17016} (\bibinfo{year}{2017}).

\bibitem{Schlickriede2017AdvMat}
\bibinfo{author}{Schlickriede, C.} \emph{et~al.}
\newblock \bibinfo{title}{Imaging through nonlinear metalens using second
  harmonic generation}.
\newblock \emph{\bibinfo{journal}{Adv. Mater.}} \textbf{\bibinfo{volume}{30}},
  \bibinfo{pages}{1703843--n/a} (\bibinfo{year}{2018}).

\bibitem{Komar2017ApplPhysLett}
\bibinfo{author}{Komar, A.} \emph{et~al.}
\newblock \bibinfo{title}{Electrically tunable all-dielectric optical
  metasurfaces based on liquid crystals}.
\newblock \emph{\bibinfo{journal}{Appl. Phys. Lett.}}
  \textbf{\bibinfo{volume}{110}}, \bibinfo{pages}{071109}
  (\bibinfo{year}{2017}).

\bibitem{Paniagua2017arXiv}
\bibinfo{author}{Paniagua-Dominguez, R.} \emph{et~al.}
\newblock \bibinfo{title}{A metalens with near-unity numerical aperture}.
\newblock \emph{\bibinfo{journal}{https://arxiv.org/abs/1705.00895}}
  (\bibinfo{year}{2017}).

\bibitem{Zhan2017SciRep}
\bibinfo{author}{Zhan, A.}, \bibinfo{author}{Colburn, S.},
  \bibinfo{author}{Dodson, C.~M.} \& \bibinfo{author}{Majumdar, A.}
\newblock \bibinfo{title}{Metasurface freeform nanophotonics}.
\newblock \emph{\bibinfo{journal}{Sci. Rep.}} \textbf{\bibinfo{volume}{7}},
  \bibinfo{pages}{1673} (\bibinfo{year}{2017}).

\bibitem{Arbabi2015NatNano}
\bibinfo{author}{Arbabi, A.}, \bibinfo{author}{Horie, Y.},
  \bibinfo{author}{Bagheri, M.} \& \bibinfo{author}{Faraon, A.}
\newblock \bibinfo{title}{Dielectric metasurfaces for complete control of phase
  and polarization with subwavelength spatial resolution and high
  transmission}.
\newblock \emph{\bibinfo{journal}{Nat. Nanotechnol.}}
  \textbf{\bibinfo{volume}{10}}, \bibinfo{pages}{937--943}
  (\bibinfo{year}{2015}).

\bibitem{Arbabi2016NatCommun}
\bibinfo{author}{Arbabi, A.} \emph{et~al.}
\newblock \bibinfo{title}{Miniature optical planar camera based on a wide-angle
  metasurface doublet corrected for monochromatic aberrations}.
\newblock \emph{\bibinfo{journal}{Nat. Commun.}} \textbf{\bibinfo{volume}{7}},
  \bibinfo{pages}{13682} (\bibinfo{year}{2016}).

\bibitem{Arbabi2017NatPhoton}
\bibinfo{author}{Arbabi, A.}, \bibinfo{author}{Arbabi, E.},
  \bibinfo{author}{Horie, Y.}, \bibinfo{author}{Kamali, S.~M.} \&
  \bibinfo{author}{Faraon, A.}
\newblock \bibinfo{title}{Planar metasurface retroreflector}.
\newblock \emph{\bibinfo{journal}{Nat. Photon.}} \textbf{\bibinfo{volume}{11}},
  \bibinfo{pages}{415--420} (\bibinfo{year}{2017}).

\bibitem{Groever2017NanoLett}
\bibinfo{author}{Groever, B.}, \bibinfo{author}{Chen, W.~T.} \&
  \bibinfo{author}{Capasso, F.}
\newblock \bibinfo{title}{Meta-lens doublet in the visible region}.
\newblock \emph{\bibinfo{journal}{Nano Lett.}} \textbf{\bibinfo{volume}{17}},
  \bibinfo{pages}{4902--4907} (\bibinfo{year}{2017}).

\bibitem{Arbabi2017arXiv}
\bibinfo{author}{Arbabi, E.} \emph{et~al.}
\newblock \bibinfo{title}{$\mathrm{MEMS}$-tunable metasurface lens}.
\newblock \emph{\bibinfo{journal}{https://arxiv.org/abs/1712.06548}}
  (\bibinfo{year}{2017}).

\bibitem{Ellenbogen2012NanoLett}
\bibinfo{author}{Ellenbogen, T.}, \bibinfo{author}{Seo, K.} \&
  \bibinfo{author}{Crozier, K.~B.}
\newblock \bibinfo{title}{Chromatic plasmonic polarizers for active visible
  color filtering and polarimetry}.
\newblock \emph{\bibinfo{journal}{Nano Lett.}} \textbf{\bibinfo{volume}{12}},
  \bibinfo{pages}{1026--1031} (\bibinfo{year}{2012}).

\bibitem{Dandan2015OptExp}
\bibinfo{author}{Wen, D.} \emph{et~al.}
\newblock \bibinfo{title}{Metasurface for characterization of the polarization
  state of light}.
\newblock \emph{\bibinfo{journal}{Opt. Express}} \textbf{\bibinfo{volume}{23}},
  \bibinfo{pages}{10272--10281} (\bibinfo{year}{2015}).

\bibitem{Khorasaninejad2015Optica}
\bibinfo{author}{Khorasaninejad, M.}, \bibinfo{author}{Zhu, W.} \&
  \bibinfo{author}{Crozier, K.~B.}
\newblock \bibinfo{title}{Efficient polarization beam splitter pixels based on
  a dielectric metasurface}.
\newblock \emph{\bibinfo{journal}{Optica}} \textbf{\bibinfo{volume}{2}},
  \bibinfo{pages}{376--382} (\bibinfo{year}{2015}).

\bibitem{Chen2016Nanotechnology}
\bibinfo{author}{Chen, W.~T.} \emph{et~al.}
\newblock \bibinfo{title}{Integrated plasmonic metasurfaces for
  spectropolarimetry}.
\newblock \emph{\bibinfo{journal}{Nanotechnology}}
  \textbf{\bibinfo{volume}{27}}, \bibinfo{pages}{224002}
  (\bibinfo{year}{2016}).

\bibitem{Mueller2016Optica}
\bibinfo{author}{Balthasar~Mueller, J.~P.}, \bibinfo{author}{Leosson, K.} \&
  \bibinfo{author}{Capasso, F.}
\newblock \bibinfo{title}{Ultracompact metasurface in-line polarimeter}.
\newblock \emph{\bibinfo{journal}{Optica}} \textbf{\bibinfo{volume}{3}},
  \bibinfo{pages}{42--47} (\bibinfo{year}{2016}).

\bibitem{Ding2017ACSPhotonics}
\bibinfo{author}{Ding, F.}, \bibinfo{author}{Pors, A.}, \bibinfo{author}{Chen,
  Y.}, \bibinfo{author}{Zenin, V.~A.} \& \bibinfo{author}{Bozhevolnyi, S.~I.}
\newblock \bibinfo{title}{Beam-size-invariant spectropolarimeters using
  gap-plasmon metasurfaces}.
\newblock \emph{\bibinfo{journal}{ACS Photonics}} \textbf{\bibinfo{volume}{4}},
  \bibinfo{pages}{943--949} (\bibinfo{year}{2017}).

\bibitem{Huard1997Polarization}
\bibinfo{author}{Huard, S.}
\newblock \emph{\bibinfo{title}{Polarization of light}},
  vol.~\bibinfo{volume}{1} (\bibinfo{publisher}{Wiley-VCH},
  \bibinfo{year}{1997}).

\bibitem{Liu2012CompPhys}
\bibinfo{author}{Liu, V.} \& \bibinfo{author}{Fan, S.}
\newblock \bibinfo{title}{S4 : A free electromagnetic solver for layered
  periodic structures}.
\newblock \emph{\bibinfo{journal}{Comput. Phys. Commun.}}
  \textbf{\bibinfo{volume}{183}}, \bibinfo{pages}{2233--2244}
  (\bibinfo{year}{2012}).

\bibitem{Arbabi2016OptExp}
\bibinfo{author}{Arbabi, E.}, \bibinfo{author}{Arbabi, A.},
  \bibinfo{author}{Kamali, S.~M.}, \bibinfo{author}{Horie, Y.} \&
  \bibinfo{author}{Faraon, A.}
\newblock \bibinfo{title}{High efficiency double-wavelength dielectric
  metasurface lenses with dichroic birefringent meta-atoms}.
\newblock \emph{\bibinfo{journal}{Opt. Express}} \textbf{\bibinfo{volume}{24}},
  \bibinfo{pages}{18468--18477} (\bibinfo{year}{2016}).

\bibitem{Arbabi2017SPIEPW}
\bibinfo{author}{Arbabi, A.} \emph{et~al.}
\newblock \bibinfo{title}{Increasing efficiency of high-$\mathrm{NA}$
  metasurface lenses}.
\newblock In \emph{\bibinfo{booktitle}{SPIE Photon. West}}, vol.
  \bibinfo{volume}{10113}, \bibinfo{pages}{101130K--1}
  (\bibinfo{publisher}{SPIE}, \bibinfo{year}{2017}).

\bibitem{Kamali2016LaserPhotonRev}
\bibinfo{author}{Kamali, S.~M.}, \bibinfo{author}{Arbabi, E.},
  \bibinfo{author}{Arbabi, A.}, \bibinfo{author}{Horie, Y.} \&
  \bibinfo{author}{Faraon, A.}
\newblock \bibinfo{title}{Highly tunable elastic dielectric metasurface
  lenses}.
\newblock \emph{\bibinfo{journal}{Laser Photon. Rev.}}
  \textbf{\bibinfo{volume}{10}}, \bibinfo{pages}{1062--1062}
  (\bibinfo{year}{2016}).

\bibitem{Sell2017NanoLett}
\bibinfo{author}{Sell, D.}, \bibinfo{author}{Yang, J.},
  \bibinfo{author}{Doshay, S.}, \bibinfo{author}{Yang, R.} \&
  \bibinfo{author}{Fan, J.~A.}
\newblock \bibinfo{title}{Large-angle, multifunctional metagratings based on
  freeform multimode geometries}.
\newblock \emph{\bibinfo{journal}{Nano Lett.}} \textbf{\bibinfo{volume}{17}},
  \bibinfo{pages}{3752--3757} (\bibinfo{year}{2017}).

\bibitem{Arbabi2017Optica}
\bibinfo{author}{Arbabi, E.}, \bibinfo{author}{Arbabi, A.},
  \bibinfo{author}{Kamali, S.~M.}, \bibinfo{author}{Horie, Y.} \&
  \bibinfo{author}{Faraon, A.}
\newblock \bibinfo{title}{Controlling the sign of chromatic dispersion in
  diffractive optics with dielectric metasurfaces}.
\newblock \emph{\bibinfo{journal}{Optica}} \textbf{\bibinfo{volume}{4}},
  \bibinfo{pages}{625--632} (\bibinfo{year}{2017}).

\bibitem{Bayer1976Patent}
\bibinfo{author}{Bayer, B.}
\newblock \bibinfo{title}{Color imaging array}.
\newblock \emph{\bibinfo{journal}{$\mathrm{US~Patent}$}}
  \bibinfo{pages}{3,971,065} (\bibinfo{year}{1976}).

\bibitem{Arbabi2016SciRep}
\bibinfo{author}{Arbabi, E.}, \bibinfo{author}{Arbabi, A.},
  \bibinfo{author}{Kamali, S.~M.}, \bibinfo{author}{Horie, Y.} \&
  \bibinfo{author}{Faraon, A.}
\newblock \bibinfo{title}{Multiwavelength metasurfaces through spatial
  multiplexing}.
\newblock \emph{\bibinfo{journal}{Sci. Rep.}} \textbf{\bibinfo{volume}{6}},
  \bibinfo{pages}{32803} (\bibinfo{year}{2016}).

\bibitem{Lin2016NanoLett}
\bibinfo{author}{Lin, D.} \emph{et~al.}
\newblock \bibinfo{title}{Photonic multitasking interleaved $\mathrm{S}$i
  nanoantenna phased array}.
\newblock \emph{\bibinfo{journal}{Nano Lett.}} \textbf{\bibinfo{volume}{16}},
  \bibinfo{pages}{7671--7676} (\bibinfo{year}{2016}).

\bibitem{Backlund2016NatPhoton}
\bibinfo{author}{Backlund, M.~P.} \emph{et~al.}
\newblock \bibinfo{title}{Removing orientation-induced localization biases in
  single-molecule microscopy using a broadband metasurface mask}.
\newblock \emph{\bibinfo{journal}{Nat. Photon.}} \textbf{\bibinfo{volume}{10}},
  \bibinfo{pages}{459--462} (\bibinfo{year}{2016}).

\end{thebibliography}

\clearpage

\clearpage

\clearpage

\clearpage

\end{document}